\documentclass[prl,a4paper,twocolumn]{revtex4}
\usepackage{graphicx,amsmath,amsfonts,latexsym,color,dcolumn,bm}
\usepackage{pdfsync}
\usepackage{hyperref} 
\usepackage{graphicx}
\usepackage[latin1]{inputenc}

\providecommand{\ket}[1]{\lvert #1 \rangle}
\providecommand{\braket}[2]{\langle #1 \vert #2 \rangle}

\newcommand{\DOIURL}[1]{http://dx.doi.org/#1} 
\newcommand{\PRL}[2]%
   {\href{\DOIURL{10.1103/PhysRevLett.#1.#2}}%
   {Phys. Rev. Lett. \textbf{#1}, #2}} 
\newcommand{\PRA}[2]%
   {\href{\DOIURL{10.1103/PhysRevA.#1.#2}}%
   {Phys. Rev. A \textbf{#1}, #2}} 
\newcommand{\PRB}[2]%
   {\href{\DOIURL{10.1103/PhysRevB.#1.#2}}%
   {Phys. Rev. B \textbf{#1}, #2}} 
\newcommand{\RevMP}[2]%
   {\href{\DOIURL{10.1103/RevModPhys.#1.#2}}%
   {Rev. Mod. Phys. \textbf{#1}, #2}} 
\newcommand{\QPH}[1]%
   {\href{http://arxiv.org/abs/quant-ph/#1}%
   {\texttt{e-print arxiv : quant-ph/#1}}} 
\newcommand{\Arxiv}[1]%
   {\href{http://arxiv.org/abs/#1}%
   {\texttt{e-print arxiv:#1}}}

\begin{document}

\title{Robust preparation and manipulation of protected qubits using time--varying Hamiltonians}

\author{Thomas Coudreau$^1$\footnote{email: thomas.coudreau@univ-paris-diderot.fr}, Beno\^it Dou\c cot$^2$, Romain Dubessy$^1$, Daria Andreoli$^1$, Perola Milman$^{1,3}$}

\address{$^1$ Laboratoire Mat\'{e}riaux et Ph\'{e}nom\`{e}nes Quantiques, CNRS UMR
  7162, Universit\'{e} Paris Diderot, 75013 Paris,
  France;\\ $^2$ Laboratoire de Physique
  Th\'{e}orique et Hautes \'{E}nergies, CNRS UMR 7589, Universit\'{e} Pierre et Marie Curie, 75005 Paris, France;\\$^3$ Institut des Sciences Moléculaires d'Orsay (ISMO), Université Paris-Sud, F-91405 Orsay, France}

\begin{abstract}
We show that it is possible to initialize and manipulate in a deterministic manner protected qubits using time varying Hamiltonians.
Taking advantage of the  symmetries of the system, we predict the effect of the noise during the initialization and manipulation.
These predictions are in good agreement with numerical  simulations.
Our study shows that the topological protection remains efficient under realistic experimental conditions.
\end{abstract}

\maketitle

\textit{Introduction.} --- Quantum error correcting codes are deemed a crucial tool for the implementation of quantum algorithms \cite{Shor95,Steane96}. 
Soon after their discovery, it was shown that some errors can be self--corrected by assigning them an energetic penalty. 
This is based on the encoding of a single logical qubit in the degenerate ground state of a composite system whose excited states are separated by a finite energy gap from the ground state subspace \cite{Bacon01,Jordan06,Weinstein07}.
Following the pioneering proposal by Kitaev \cite{Kitaev97}, it has been shown that when this scheme is implemented using specific Hamiltonians with collective symmetries, it can reduce exponentially the influence of decoherence on logical qubits and quantum gates \cite{Denis02,Ioffe02,Doucot05,Bacon06,Milman07}.

It is usually recognized that logical qubits should verify the so-called di Vincenzo criteria \cite{Divincenzo01}. 
In particular, it is necessary to be able to initialize to a fiducial state and perform single and two qubits rotations on them. 
This initialization is difficult because the protected states are highly entangled states which are thus not reachable from simple, separable states using local operations. 
Also, for a protected qubit, there is an apparent conflict between the protection and the ability to manipulate it. 
We show here how to overcome these difficulties using time varying Hamiltonians with well--defined symmetries. 
Using both high--order perturbation theory and numerical simulations, we also take into account explicitly the effect of random noise during all these processes showing that the qubits remain protected during these operations. 

%
%

{\it Protection Hamiltonian.} --- In Refs.\cite{Kitaev97,Denis02,Ioffe02,Doucot05,Bacon06,Milman07}, well--designed Hamiltonians acting on an array of $N\times N$ two--level systems were introduced and shown to yield degenerate quantum states largely immune to decoherence.
The protection relies on the existence of a set of symmetries, which are collective operators acting on rows or columns independently, \emph{i.e.} $\hat P_i = \prod_{j=1}^N \hat\sigma_{i,j}^y, \hat Q_j = \prod_{i=1}^N \hat\sigma_{i,j}^x$ where $\hat \sigma_{i,j}^{\alpha}$ is the standard Pauli matrix in the $\alpha$ direction in spin--space acting on the two-level system located at the intersection of row $i$ and column $j$.
They obey the following rules : $[\hat P_i,\hat P_{i'}]=[\hat Q_j,\hat Q_{j'}]= 0 = \{ \hat P_i, \hat Q_j\}$, $\hat P_i^2 = 1 = \hat Q_j ^2$.
The symmetries enforce the two--fold degeneracy of all eigenstates, including the ground states $\{\ket{0_L}, \ket{1_L}\}$.
It is convenient to choose these states so that $\hat P_i$ acts as a logical Pauli operator denoted $\hat \tau^z$ while $\hat Q_j$ acts as $\hat \tau^x$. 
These states are complex entangled states composed of arrays of $N^2$ two-level systems that cannot be prepared nor manipulated easily.
They are separated from the other eigenstates by a finite energy gap $\Delta$, whose exact value depends on the specific model and on the system size. lines
More quantitatively, the degeneracy lifting when the system is subjected to a static noise of amplitude $h$ independently acting on $N$  of its constituents scales as $h^N/\Delta^{N-1}$, which decreases exponentially with $N$ provided $h$ is smaller than $\Delta$.
Accordingly, the coherence time increases exponentially with $N$.

In the physically relevant case of a dynamical noise, this exponential improvement holds only below a critical size $N^\star$ which can roughly be estimated as the ratio between the energy gap $\Delta$ and the typical energy scale ($k_B T$ for a noise source in thermal equilibrium) which the environment can provide to the system in a single elementary process.
The existence of this finite $N^\star$ is consistent with the notion that topological order in 2D systems disappears at finite temperature in the thermodynamical limit \cite{Alicki06,Castelnovo07,Ortiz08}. 
However, for the purpose of quantum computation, we are interested in small systems initially prepared in their ground states and time scales much shorter than the thermalization times.

To illustrate numerically how one can circumvent the difficulties of initializing and manipulating such protected states, we will use the long--range Hamiltonian
\begin{equation}
\hat H_{0}=-J_{x}\sum_{i}^{N}\left( \sum_{j}^{N} \hat\sigma _{i,j}^{x}\right)
^{2}-J_{y}\sum_{j}^{N}\left( \sum_{i}^{N}\hat\sigma _{i,j}^{y}\right)
^{2}
\label{eq:infinite}
\end{equation}
where $J_x$ and $J_y$ are numerical constants and which can be realized either using trapped ions \cite{Milman07} or superconducting circuits \cite{Xue08a}.

\begin{figure}[h]
    \centering
    \includegraphics[width=.75\columnwidth]{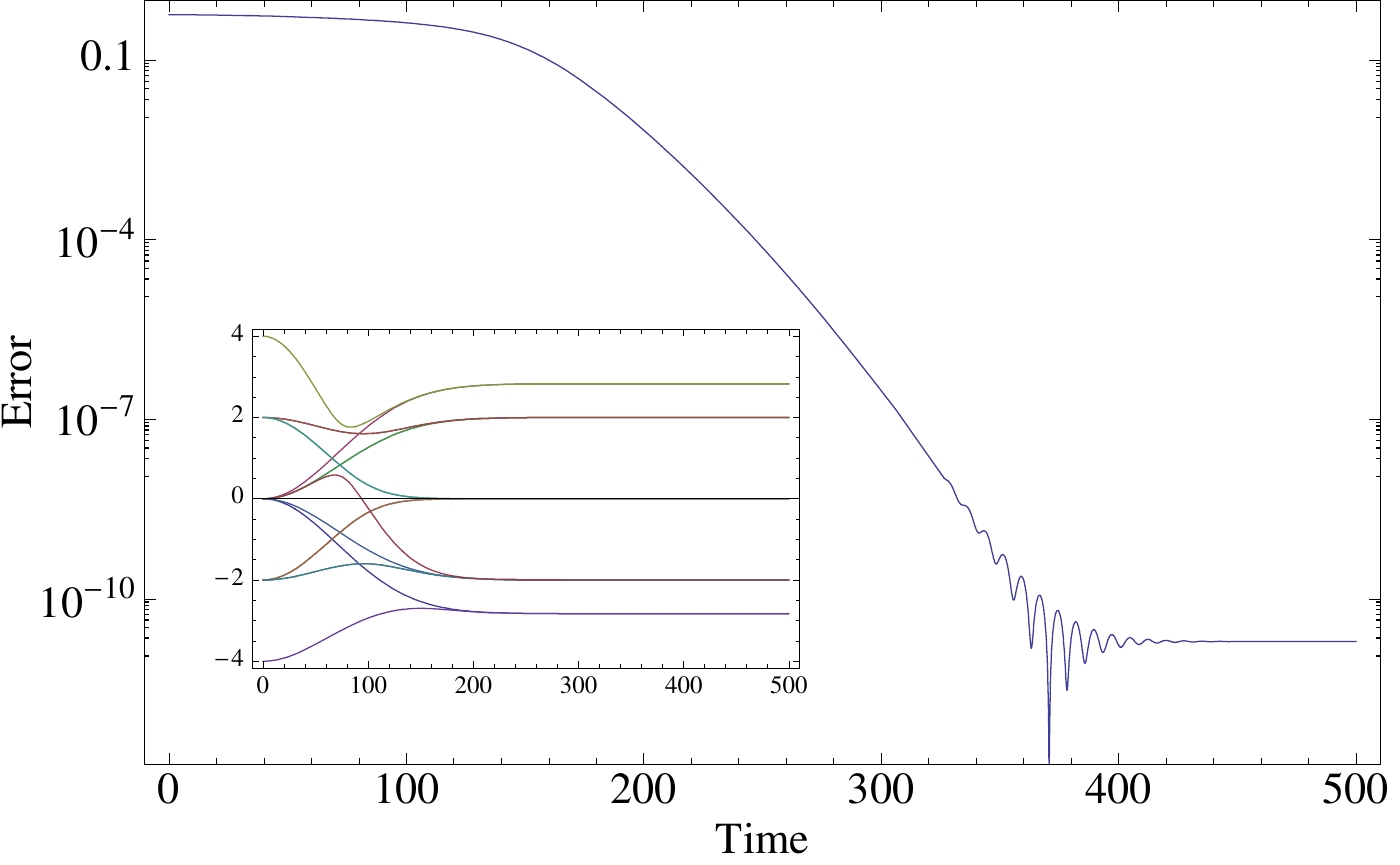}
        \includegraphics[width=.75\columnwidth]{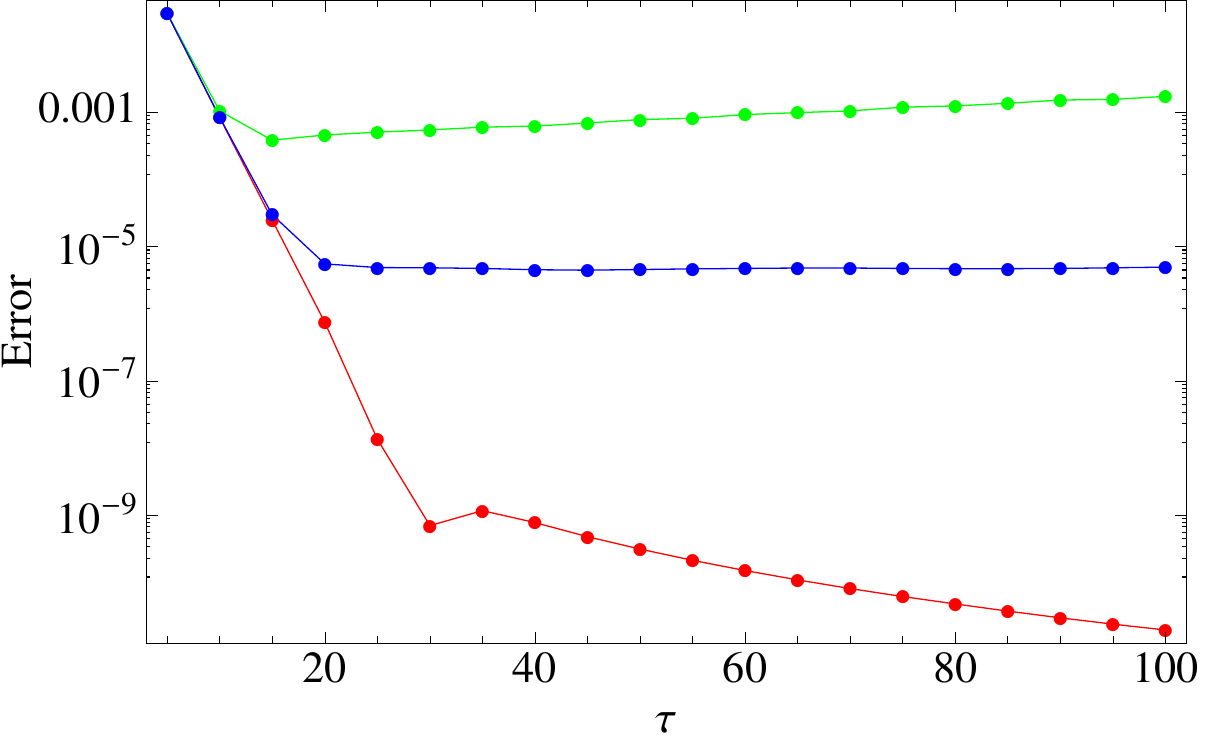}
    \caption{(Color online) Top: Example of the state evolution as a
      function of time without noise in 2$\times$2 case and for
      $\tau=100$ which is large compared to $\Delta^{-1} = (\sqrt 2 +1)/2$. 
      Inset : Evolution of the energy levels (normalized to $J=J_x=J_y$) as a function of time: at $t=0$ the energy levels are those of $\hat S^y$ and close to those of $\hat H_0$ at $t =\tau$. Bottom : Error as a function of time for a static noise applied on all qubits on a 2 $\times$ 2. Noise amplitude 10$^{-2}$ (squares) and 10$^{-3}$ (circles)}
    \label{fig:initialization}
  \end{figure}

{\it Initialization.} --- We now turn to the state initialization. 
In the same spirit as \cite{Hamma08}, we proceed by applying a static field that acts independently on each two-level system.
Such a local field is then adiabatically turned off from $t=0$ to $t=T$ while the protecting Hamiltonian is turned on during the same interval, in a process described by the time dependent Hamiltonian
  $\hat H_{t}=f(t) \hat H_{0} + (1-f(t)) \hat S^y,$ 
where $f(t)$ is a slowly varying function such that $f(0)=0$ and $f(t) \stackrel{t\to + \infty}{\longrightarrow} 1$ with a characteristic time $\tau$, and $\hat S^{y}=\sum_{i,j}^N \sigma_{i,j}^y$ describes the effect of the static field. 
The system can be prepared through single site operations in the (ferromagnetic) ground state of $\hat S^y$.
This state is an eigenstate of $\hat P_i$ with the eigenvalue 1 which ensures that, after the evolution, the final state is the protected eigenstate with the same eigenvalues, $\ket{0_L}$.
In such a state preparation scheme, the system only reaches asymptotically the decoherence protected state. 
The time evolution of the instantaneous energy levels is indicated in Fig.~\ref{fig:initialization}~(inset): at $t=0$ the energy levels are those of $\hat S^y$ and close to those of $\hat H_0$ at $t=5\tau$ (where we have taken $f(t)=\exp(-t^2/\tau^2)$).
The effect of this procedure on the logical state is shown in Fig.~\ref{fig:initialization}, top where we have plotted the error $1-F(t)$ where $F(t)$ is the fidelity $F(t)=|\braket{0_L}{\psi(t)}|^2$ of the prepared state, with $\ket{\psi(t)}$ the quantum state of the system at time $t$.
This figure shows that for $\tau = 100$, the final state is extremely close to the target state(error of the order of 10$^{-10}$). Indeed, the error after a time $t=5\tau$ diminishes as $\tau$ increases and the error is below 10$^{-5}$ for $\tau \geq 20$ as shown in Fig.~\ref{fig:initialization}, bottom.

However, in a real experiment, during the whole evolution, the system is subjected to noise that will render state preparation imperfect.
An important point is to evaluate the effect of noise on the fidelity.
Simulations show that the preparation scheme is quite robust to noise. 
We have plotted in Fig.~\ref{fig:initialization} the error for different noise amplitudes as a function of $\tau$.
We consider here a static noise with a random orientation applied on each of the individual two-level systems.
This figure shows that even for realistic error rates, on the order of 0.01$J$, there exists some values of $\tau$ for which the error will remain smaller than 10$^{-3}$ (2$\times$2 case).

Our results are important for the realization of experimental tests of protected qubits as they show that, under realistic experimental conditions, errors will not destroy the quality of the prepared state.
Let us now turn our attention to the manipulation of a single protected qubit.

{\it Manipulation.}--- We address the possibility of manipulating the ground states of the protecting Hamiltonian using local operations.\\
We remark that, due to the symmetry of the system, an operator acts within the logical qubit space if it either preserves or changes into their opposites \emph{simultaneously all} the eigenvalues of the symmetry operators.
This condition will be denoted in the following Condition 1.

Firstly, we study the noiseless case. We thus consider a perturbation Hamiltonian of the form $\hat{\mathcal S}^{u}= \sum_{i,j} \hat \sigma^{u}_{i,j}$ where $u= x$ or $y$. 
This Hamiltonian is assumed to be applied using a smooth time-varying gate function so that the total Hamiltonian reads $\hat H_{t} = \hat H_{0} + g^u(t) \hat {\mathcal S}^u$.
The maximal value attained by $g^u(t)$ , $g_{max}^u$ remains small compared to the energy gap of $\hat H_{0}$ so that no level crossing occur. 
Furthermore, the typical variation time is chosen much smaller than the inverse of the gap frequency so that the transformation can be considered as adiabatic within the excited manifold (it will naturally not be adiabatic within the protected state where the energy difference is initially zero).
According to the quantum adiabatic theorem, these conditions are sufficient to ensure that the final state will remain within the protected subspace ($\ket{0_L},\ket{1_L}$).
Thus, the unitary operator $\hat U$ describing the transformation from initial to final state can be decomposed over the logical Pauli operators, denoted $\hat \tau^{x,y,z}$, and the two--dimensional identity matrix, $\hat{\mathbb I}_2$
\begin{equation}\label{U}
\hat U= \alpha_1 \hat{\mathbb I}_2 + \sum_{i=x, y, z} \alpha_x \hat \tau^x
\end{equation}
In order to study the effect of the manipulation, one can apply high order perturbation theory and decompose the result according to \eqref{U}.
At the $M^\text{th}$ order, one obtains a sum of terms of the form $\hat H_M=\hat \sigma ^u_{i_1,j_1} \hat \sigma ^u_{i_2,j_2} \dots \hat \sigma ^u_{i_M,j_M}$ ($u$ is an arbitrary direction).
Due to the adiabatic theorem, $\hat H_M$ will act within the logical qubit space so that it must verify Condition (1) to have a non trivial effect.
Two pairs of exclusive possibilities can then be considered:
\\ 1. the first pair concerning the $\hat P_i$ operators (whose eigenvalues are denoted $p_i$)
  \\ \indent (a) $\hat H_M$ conserves all $p_i$: in this case, $\hat H_M$ commutes with the symmetry operators $\hat P_i$ and thus with the logical operator $\hat \tau^z$. It must then induce a term of the form $a \hat{\mathbb I}_2 + b \hat \tau^z$ where, here and in the following, $a$ and $b$ are constants depending on the exact shape of $g^u$;
  \\ \indent (b) $\hat H_M$ changes all $p_i$ in their opposites: in this case, $\hat H_M$ anti--commutes with the symmetry operators $\hat P_i$ and thus with the logical operator $\hat \tau^z$. It must then induce a term  of the form $a \hat \tau^x + b \hat \tau^y$;
\\ 2. the second pair concerning the $\hat Q_j$ operators (whose eigenvalues are denoted $q_j$)
  \\ \indent (c) $\hat H_M$ conserves all $q_j$: in this case, $\hat H_M$ commutes with the symmetry operators $\hat Q_j$ and thus with  the logical operator$\hat \tau^x$. It must then induce a term of the form $a \hat{\mathbb I}_2 + b \hat \tau^x$;
  \\ \indent (d) $\hat H_M$ changes all $q_j$ in their opposites: in this case, $\hat H_M$ anti--commutes with the symmetry operators $\hat Q_j$ and thus with the logical operator $\hat \tau^x$. It must then induce a term of the form $a \hat \tau^y + b \hat \tau^z$.

For a given Hamiltonian $H_M$, only one term will remain for the two pairs of conditions which can be summarized in Tab.~(\ref{tab:1}).
We see that, for each of the four conditions, one term and only one will survive which simplifies greatly the analysis.
\begin{table}[h]
  \begin{tabular}{ccc}
  &
  \begin{tabular}{c}
Condition (c) \\($q_j \to q_j, \forall j$)
\end{tabular}
&   
  \begin{tabular}{c} Condition (d) \\($q_j \to - q_j, \forall j$) \end{tabular}\\\hline\hline
  \begin{tabular}{c} Condition (a) \\($p_i \to p_i, \forall i$) \end{tabular}& $\mathbb I$ & $\hat \tau^z$\\
  \begin{tabular}{c}
Condition (b)\\ ($p_i \to -p_i, \forall i$) \end{tabular}& $\hat \tau^x$ & $\hat \tau^y$\\
  \end{tabular}
\caption{\label{tab:1}Effective operator induced by the manipulation as a function of its effect on the ground--state quantum numbers.}
\end{table}

As discussed above, high order perturbation theory yields a sum of terms of the form $\hat H_M, M \in \mathbb{N}.$ 
Each of these terms will have an effect summarized in Tab.~\eqref{tab:1} and we have to consider the dominant terms in the sum.
Let us first consider the case of a manipulation along $y$.
At any order of perturbation theory,  $\hat{\mathcal S}^y$ will conserve the $p_i$ quantum numbers, all terms appearing in the perturbation theory development are thus in the first line of Tab.~(\ref{tab:1}).
For $M<N$, there is at least one $q_j$ quantum number which is preserved so that  Condition (1) can not be fulfilled: the manipulation in this case has no effect, in good agreement with the idea of protection.
The first, non--zero term will appear when all columns are touched at least once by the manipulation (condition (d)), that is $M=N$.
In this case, the effect will be on the order of $\left(\frac{g_{max}^u}{\Delta}\right)^N \frac{\Delta t}{\hbar} \tau^z$ where $t$ is the duration of the manipulation.
A similar reasoning can be made for a manipulation along $x$ which preserves the $q_j$ quantum numbers: again, the first, non--zero term will appear when all lines are touched at least once by the manipulation, that is $M=N$.
In this case, the effect will be of the order of $\left(\frac{g_{max}^u}{\Delta}\right)^N \frac{\Delta t}{\hbar}\tau^x$.
Finally, combining these two types of operations, any arbitrary rotation on the protected qubit can be performed, a key operation toward the use of such systems for quantum calculation purposes.

Let us now tackle the more difficult question of the resistance of this procedure to noise.
We consider here an arbitrary static noise which is expressed as a sum of single spin terms.
In practice, one will overestimate the effect noise by taking the dominant noise assumed to be along $v$ with amplitude $f^v$.
The total Hamiltonian now reads $ \hat H = \hat H_0 + g(t) \hat {\mathcal S}^u + f \hat {\mathcal S}^v$.

At the $M^\text{th}$ order of perturbation theory, one is left with a sum of terms of the form $\hat H_M=\left(\hat \sigma ^u_{i_1,j_1} \hat \sigma ^u_{i_2,j_2} \dots \hat \sigma ^u_{i_m,j_m}\right) \left(\hat \sigma ^v_{i_1,j_1} \hat \sigma ^v_{i_2,j_2} \dots \hat \sigma ^v_{i_n,j_n}\right)$ where the first term comes from the manipulation term and the second from the noise term. 
We have naturally $m+n=M$. 
In the case where manipulation and noise are parallel ($u=x =v$ or $u=y=v$), the two terms add up to yield a term of the form $(g^u_{max}+f)^N \hat \tau^{x,z}$.
In this case, the rotation axis is not modified but the rotation angle is affected to first order in $f/g^u_{max}$, without any gain from the protection.

We now turn our attention to a case where manipulation and noise are not parallel, for instance $u=y$ and $v=x$.
The two terms affect respectively only the  columns (manipulation) or the lines (noise).
The previous result can be used: one can have a change of sign of a family of quantum numbers only if the corresponding power ($m$ for the manipulation and thus the columns and $n$ for the noise and thus the lines) is larger than or equal to $N$: the smallest order for noise is thus either 2 
 (when there is no sign change of the $p_i$) or $N$ (when all the $p_i$ change sign) while the smallest order for the manipulation is either 0  (when there is no sign change of the $q_j$) or $N$ (when all the $q_j$ change sign).
Finally, these terms must be compared with the effect of the manipulation (proportional to $\left(\frac{g_{max}^u}{\Delta}\right)^N \frac{\Delta t}{\hbar}$) as it is the ratio which is relevant and the corresponding results are detailed in Tab.~(\ref{tab:3}).
\begin{table}[t]
  \begin{tabular}{ccc}
  &
  \begin{tabular}{c}
Condition (c) \\($q_j \to q_j, \forall j$)
\end{tabular}
&   
  \begin{tabular}{c} Condition (d) \\($q_j \to - q_j, \forall j$) \end{tabular}\\\hline\hline
  \begin{tabular}{c} Condition (a) \\($p_i \to p_i, \forall i$) \end{tabular}& not relevant & $\frac{ f^{2}}{\Delta^{2}} \hat \tau^z$\\
  \begin{tabular}{c}
Condition (b)\\ ($p_i \to -p_i, \forall i$) \end{tabular}& $ \frac{f^N}{(g^x_{max})^N} \hat \tau^x$ & $\frac{f^{N}}{\Delta^{N}}\hat \tau^y$\\
  \end{tabular}
\caption{\label{tab:3}Dominant term induced by the manipulation and noise terms as a function of its effect on the quantum numbers where amplitudes are normalized by the effect of the manipulation in the absence of noise.}
\end{table}
Similar results can be obtained for the other cases where noise is not parallel to the manipulation.
We have thus demonstrated that manipulation of the logical qubit state can be performed in a deterministic manner and that this procedure is relatively resistant to noise. 
In the worst case (noise parallel to the manipulation), the noise sensitivity is the same as in the absence of protection, while for the other cases it varies at least quadratically with the perturbation strength, and the deviations of the rotation axis in qubit space are exponentially small in $N$.

In order to confirm these results, we used numerical simulations performed on $2 \times 2$ arrays.
We first checked the effect of a manipulation along $x$ in the absence of noise.
As expected, the coefficients along $\alpha_y$ and $\alpha_z$ are close to zero (typically smaller than $10^{-9}$.), showing that the manipulation is a pure rotation along $z$ of the logical qubit.
$\alpha_1$ and $\alpha_x$ are, also with a good precision, sinusoidal functions of the manipulation amplitude, $g^x$.
This can be seen as Rabi--like oscillations between the two--qubit states.

\begin{figure}[t]
\includegraphics[width=.75\columnwidth]{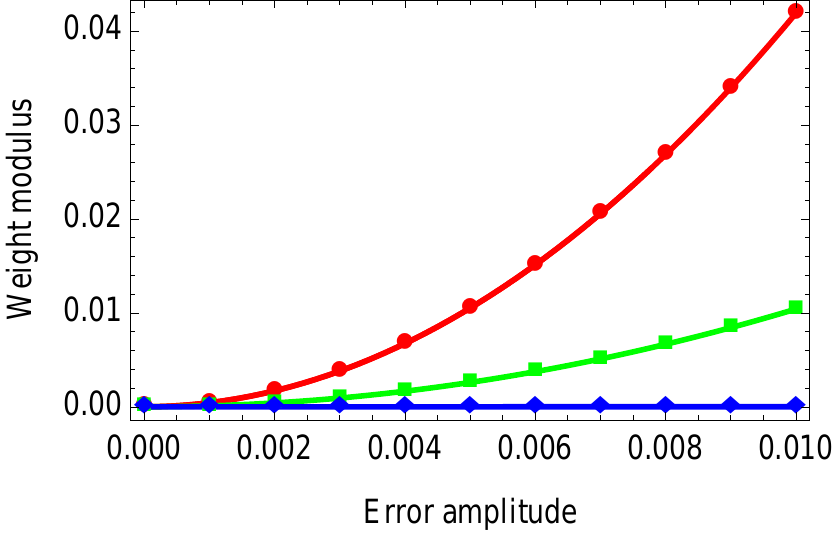}
\caption{Deviations of $|\alpha_x|$ (red circles), $|\alpha_y|$ (green squares) and $|\alpha_z|$ (blue rhombi) away from the noiseless manipulation case as a function of the error amplitude $f$ for a $2\times 2$ array.
The manipulation amplitude is $g^y_{max}=0.304$ corresponding to a $\frac \pi 8$ phase--shift. \label{fig:amplit-manip}}
\end{figure}

Concerning the effect of the noise, the effect of the protection is most apparent for a noise perpendicular to the manipulation axis.
Although we have explored numerically all combinations of directions for the manipulation and noise, we discuss as above the specific case $u=y$ and $v=x$ (Fig.~\ref{fig:amplit-manip}). 

Without noise, the manipulation should then be equivalent to a perfect dephasing.
We verify that the deviations of $\alpha_x$ and $\alpha_y$ away from a noiseless manipulation have a quadratic variation with the noise amplitude as expected since $N=2$ (see Tab.~\eqref{tab:3}, bottom row).
Note that the deviation of $\alpha_z$ also varies quadratically even though it is several orders of magnitude smaller.

In any physical implementation, the binary interactions between spins in the protected Hamiltonians are likely to be subjected to residual fluctuations, so that they become space and time dependent.
In contrast with early proposals of self--correcting qubits \cite{Bacon01,Jordan06,Weinstein07}, these fluctuations do not lift the ground state degeneracy and do not induce decoherence because they preserve the non--local symmetries $P_i$ and $Q_j$.
This remains valid as long as the gap remains open, that is as long as the fluctuations in the coupling are small in relative value.
During the manipulation, the effect of these fluctuations can be analyzed as discussed above.
It appears immediately that they have no effect on the rotation axis.
However, as in the case of a single spin noise parallel to the manipulation axis, the rotation angle is affected by an amount proportional to the noise amplitude.

\textit{Conclusion.} --- 
Using high order perturbation theory and exploiting the large number of symmetries present in Hamiltonians yielding protected qubits, we have shown that these qubits can be initialized and manipulated in a noise resistant manner.
This shows that there is no contradiction between the concept of protection and the ability to manipulate a qubit.

\textit{Acknowledgements.}--- B.D. acknowledges funding from ANR 06-BLAN-0218-01. We wish to thank Lev Ioffe, Otfried Gühne and Markus Grassl for stimulating discussions.

\end{document}